\documentclass{PoS}

\newcommand{\RR}{{\kern+.25em\sf{R}\kern-.78em\sf{I} \kern+.78em\kern-.25em}}
\newcommand{\NN}{{\kern+.25em\sf{N}\kern-.78em\sf{I} \kern+.78em\kern-.25em}}
\newcommand{\CC}{{\kern+.25em\sf{C}\kern-.45em\sf{{\small{I}}} \kern+.45em\kern-.25em}}

\include{epsfig}

\newcommand{\be}{\begin{equation}}
\newcommand{\ee}{\end{equation}}
\newcommand{\bea}{\begin{eqnarray}}
\newcommand{\eea}{\end{eqnarray}} 

\newcommand{\la}{\langle}
\newcommand{\ra}{\rangle}

\title{Microscopic Dirac Spectrum in a 2d Gauge Theory with 
Zero Chiral Condensate
\thanks{This work was supported by the Mexican {\it Consejo Nacional 
de Ciencia y Tecnolog\'{\i}a} (CONACyT) through project 155905/10 
``F\'{\i}sica de Part\'{\i}culas por medio de Simulaciones Num\'{e}ricas'', 
and by the {\it Croatian Ministry of Science, Education and Sports,} 
project No. 0160013. \newline \indent
We thank Poul Damgaard, Stephan D\"{u}rr,
Philippe de Forcrand, James Hetrick, Christian Hoelbling, Tamas Kov\'{a}cs 
and Andrei Smilga for helpful comments.}}

\ShortTitle{Dirac Spectrum at Zero Chiral Condensate}

\author{\speaker{Wolfgang Bietenholz}$^{\rm \ a}$,
Ivan Hip$^{\rm \ b}$ and David Landa-Marb\'{a}n$^{\rm \ a}$
\\
\ \\
\ \\
$^{\rm \ a}$ Insituto de Ciencias Nucleares, Universidad Nacional 
Aut\'{o}noma de M\'{e}xico \\
~~~~A.P.\ 70-543, C.P.\ 04510 Distrito Federal, Mexico \vspace*{2mm} \\
\ $^{\rm b}$ Faculty of Geotechnical Engineering, University of Zagreb \\
~~~~ Hallerova aleja 7, 42000 Vara\v{z}din, Croatia \\
\ \\
E-mail: \email{wolbi@nucleares.unam.mx, ivan.hip@gfv.hr, \\ 
\ \ \ \ \ \ necalanda@hotmail.com} \\ }

\abstract{Fermionic theories with a vanishing chiral condensate 
(in the chiral limit) have recently attracted considerable
interest; in particular variants of multi-flavour QCD are
candidates for this behaviour. Here we consider the 2-flavour
Schwinger model as a simple theory with this property. 
Based on simulations with light dynamical overlap fermions,
we test the hypothesis that in such models the low lying Dirac 
eigenvalues could be decorrelated. That has been observed in 
4d Yang-Mills theories at high temperature, but it cannot be 
confirmed for the 2-flavour Schwinger model.
We also discuss subtleties in the evaluation of the mass 
anomalous dimension and its IR extrapolation.}

\FullConference{31st International Symposium on Lattice Field Theory 
- LATTICE 2013\\
		July 29 - August 3, 2013\\
		Mainz, Germany}

\begin{document}

\section{Chiral symmetry and low lying Dirac eigenvalues}
\vspace*{-1mm}

For fermionic models, the chiral condensate $\Sigma = - \langle
\bar \psi \psi \rangle$ represents the order parameter, which
indicates if chiral symmetry is intact or broken. The latter is
the case at finite fermion mass $m$, but in the chiral limit
$m \to 0$ both scenarios are possible:
\begin{itemize}

\item The chiral symmetry breaking could persist, 
$\Sigma (m=0) > 0$. This is the familiar situation of
low temperature QCD, where chiral flavour symmetry breaks 
spontaneously at zero quark masses.

\item The scenario with $\Sigma (m=0) = 0$ occurs 
for instance in high temperature QCD. At low temperature,
multi-flavour variants of QCD are candidates for IR conformal
theories with this property.
However, the question whether or not IR conformality sets in
for instance at $N_{f} = 8$ or $12$ flavours is highly 
controversial; for a review, see Ref.\ \cite{DDebbio}.

Here the low lying Dirac eigenvalues $\lambda$ do not form a
Banks-Casher plateau. Instead, an obvious ansatz for their
density $\rho (\lambda )$ near zero is a power law
(with constants $c$, $\alpha$)
\be  \label{caform}
\rho (\lambda ) = c V | \lambda |^{\alpha} \ , \quad
\alpha = 1/ \delta \ .
\ee
In the following $V$ represents the space-time volume. On the 
other hand, in the case of high temperature, $V$ means only the
spatial volume, since there are no small (non-vanishing)
Dirac eigenvalues in the direction of Euclidean time.

\end{itemize}

The second scenario is also expected for the Schwinger model (QED
in two dimensions) with $N_{f} \geq 2$ flavours. In the continuum and 
large volume, A.\ Smilga derived the analytic result \cite{Smilga}
\be  \label{Smilgadelta}
\Sigma (m) \propto m^{1/\delta} \ , \quad
\delta = \frac{N_{f}+1}{N_{f}-1} \ .
\ee
Here we focus on $N_{f}=2$.

In the framework of this scenario, there is an interesting conjecture
that the low lying eigenvalues could be decorrelated and
thus follow a Poisson distribution. Based on this assumption,
T.G.\ Kov\'{a}cs derived an explicit formula for the densities of the
leading eigenvalues \cite{Tamas}. The iteration of his formula 
yields, for the $n^{\rm th}$ (non-zero) eigenvalue, the density
\be  \label{koveq}
\rho_{n}(\lambda ) = \frac{(cV)^{n}}{(n-1)! \ (\alpha + 1)^{n-1}}
\ \lambda^{n ( \alpha + 1) -1 } \ \exp \Big( - \frac{cV}{\alpha +1} 
\lambda^{\alpha + 1} \Big) \ .
\ee

\section{The Schwinger model with $N_{f}=2$ light fermions}
\vspace*{-1mm}

A refined analysis \cite{HHI} revealed that the critical exponent 
$\delta$ of the $N_{f}=2$ Schwinger model --- in an $L \times L$
volume, with gauge coupling $g$ --- actually depends on 
the Hetrick-Hosotani-Iso (HHI) parameter $l$,
\be  \label{HHIpar}
l := \frac{m}{\pi^{1/4}} \sqrt{2 L^{3} g} \quad
\left\{ \begin{array}{cccc}
\gg 1 &&& : \quad \delta = 3 \\
\ll 1 &&& : \quad \delta = 1 
\end{array} \right. \quad .
\ee
Hence eq.\ (\ref{Smilgadelta}) refers to the limit $l \gg 1$, 
whereas $l \ll 1$ corresponds to the spectrum of free fermions,
with $\rho (\lambda ) \propto \lambda^{d-1}$ \cite{LeuSmi}.

If one performs lattice simulations, the parameter $l$
is between these extreme cases, and it is difficult to
attain the behaviour of the $l \gg 1$ limit --- to the
best of our knowledge, such results have not appeared yet.

We simulated this model with dynamical chiral fermions
(more precisely: with overlap hypercube fermions \cite{ovHF,WBIH})
with two flavours of degenerate mass $m$, and $\beta = 1/g^{2}
=5$ \cite{BHSV}. This yields a mean plaquette value close
to $0.9$, which implies that lattice artifacts are small.
On the other hand, finite size effects may be significant in our
$L\times L$ volumes, $L=16 \dots 32$.
Table \ref{HHItab} displays the corresponding HHI parameter 
for the two fermion masses and the smallest and largest size $L$
that we are going to consider.
\begin{table}[h!]
\vspace*{-3mm}
\centering
\begin{tabular}{|c||c|c|}
\hline
 & $L=16$ & $L=32$ \\
\hline
\hline
$m=0.01$ & $l = 0.455$ & $l = 1.286$ \\ 
$m=0.06$ & $l = 2.728$ & $l = 7.715$ \\
\hline
\end{tabular}
\caption{The HHI parameter $l$ of eq.\ (\protect\ref{HHIpar}) for two 
fermion masses $m$ and lattices sizes $L$, at $\beta =5$.}
\label{HHItab}
\end{table}

\vspace*{-7mm}
\section{Decorrelation of small Dirac eigenvalues ?}
\vspace*{-1mm}

\begin{figure}[h!]
\begin{center}
\includegraphics[angle=0,width=.49\linewidth]{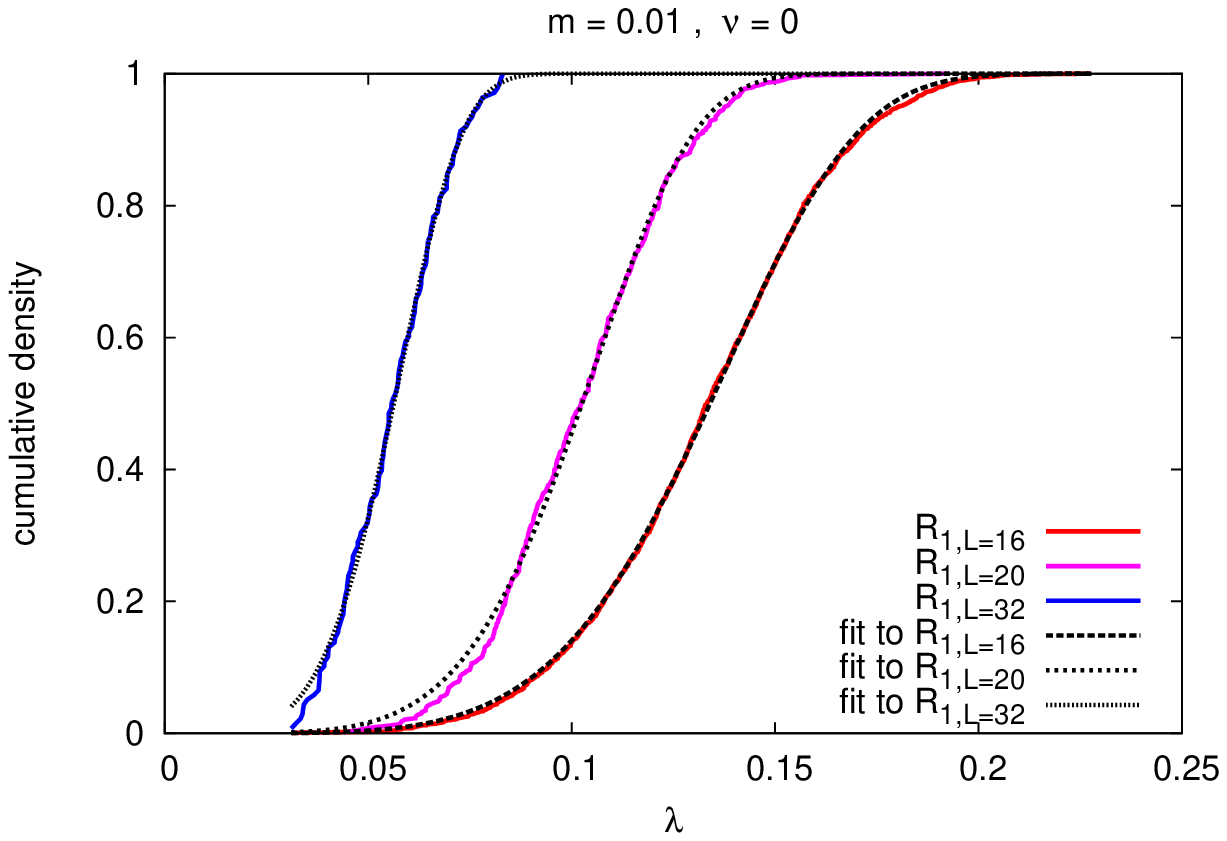} \vspace*{-3mm}\\
\includegraphics[angle=0,width=.49\linewidth]{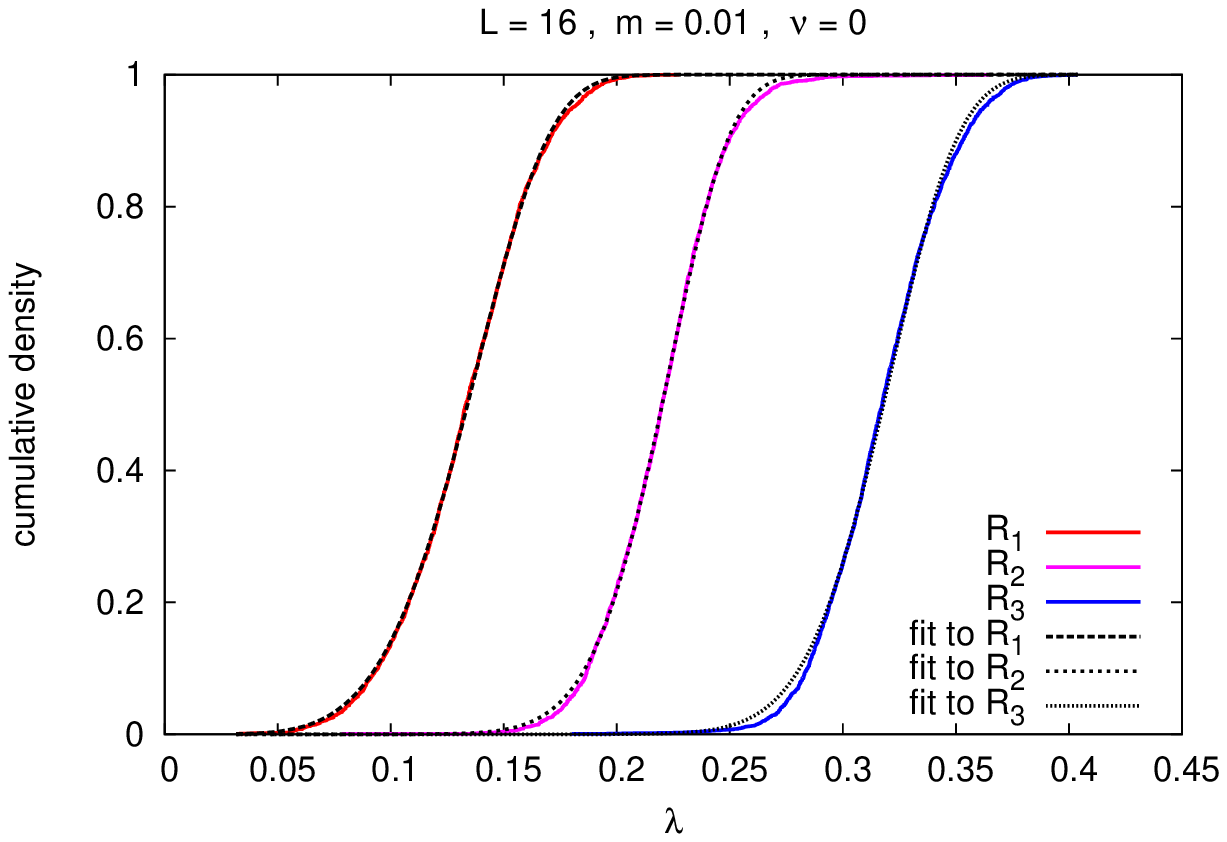}
\includegraphics[angle=0,width=.49\linewidth]{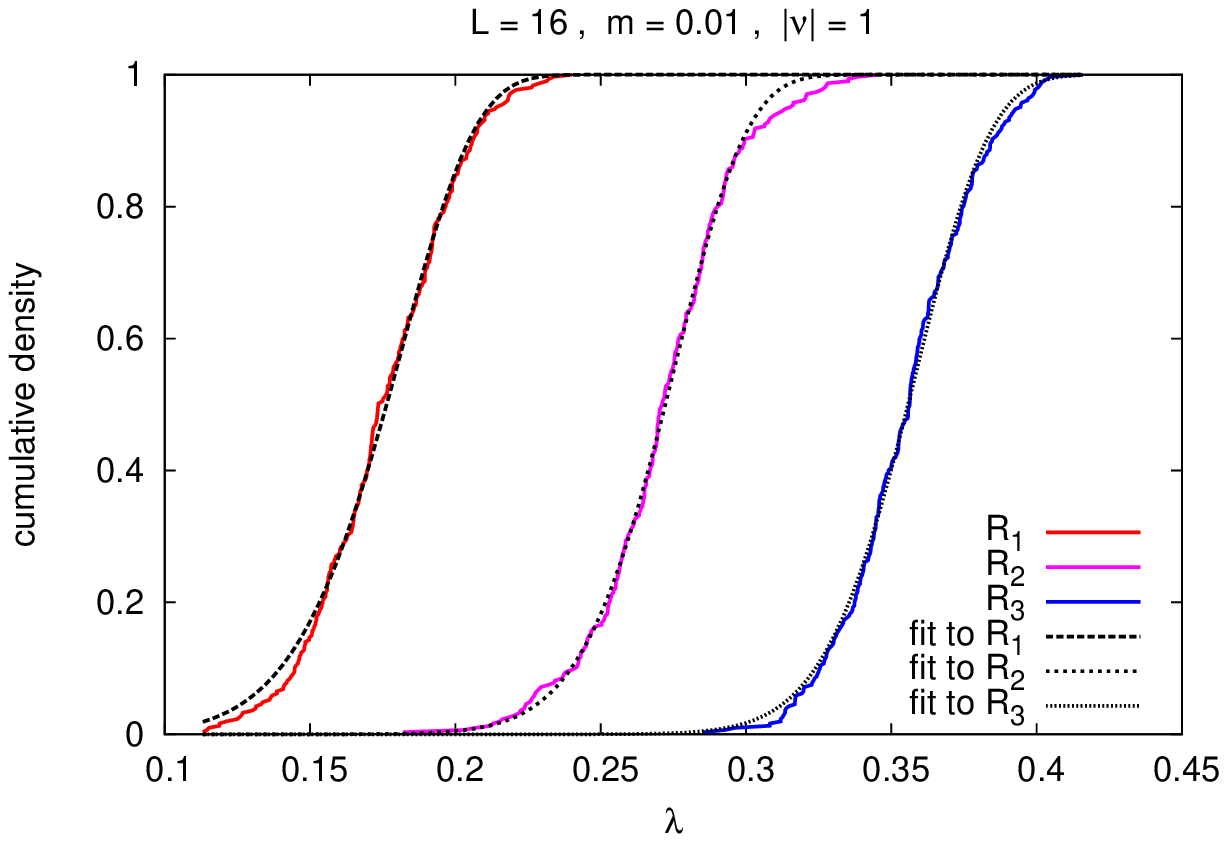}
\end{center}
\caption{Cumulative Dirac eigenvalue densities at
$m=0.01$, and the fitted Kov\'{a}cs distributions (\protect\ref{koveq}).
We show above the leading eigenvalue in the topologically
neutral sector ($\nu =0$) and lattice sizes $L=16,\ 20,\ 32$. Below
we refer to $L=16$ and include the leading three non-zero eigenvalues
in the sectors $| \nu | = 0$ and $1$.}
\label{cumdens3}
\end{figure}

The overlap Dirac operator that we used is constructed from an RG 
improved {\em hypercubic kernel} with a mass parameter $-1$ (in 
lattice units) \cite{WBIH,BHSV}. 
Thus the (massless) Dirac spectrum is located on a unit circle
in the complex plane with centre $1$. For comparison with the
continuum formula, we map the eigenvalues stereographically
onto a line, as described in Ref.\ \cite{WBStanislav}. 
In order to test the compatibility with the density (\ref{koveq}), 
we consider \cite{LMBH} the corresponding cumulative density 
(in contrast to a histogram, this does not require the choice of 
an bin size),
\be
R_{n} (\lambda) = \int_{0}^{\lambda} d \lambda' \ \rho_{n} (\lambda ') \ ,
\ee
for a fixed size $L$ and mass $m$, in some topological
sector (the topological charge is identified with the fermion 
index $\nu$). We fit the Poisson behaviour (\ref{koveq}) to
the measured cumulative densities by tuning the 
free parameters $c$ and $\alpha$. 
Figure \ref{cumdens3} shows that these fits work very well.

\begin{figure}[h!]
\begin{center}
\includegraphics[angle=0,width=.49\linewidth]{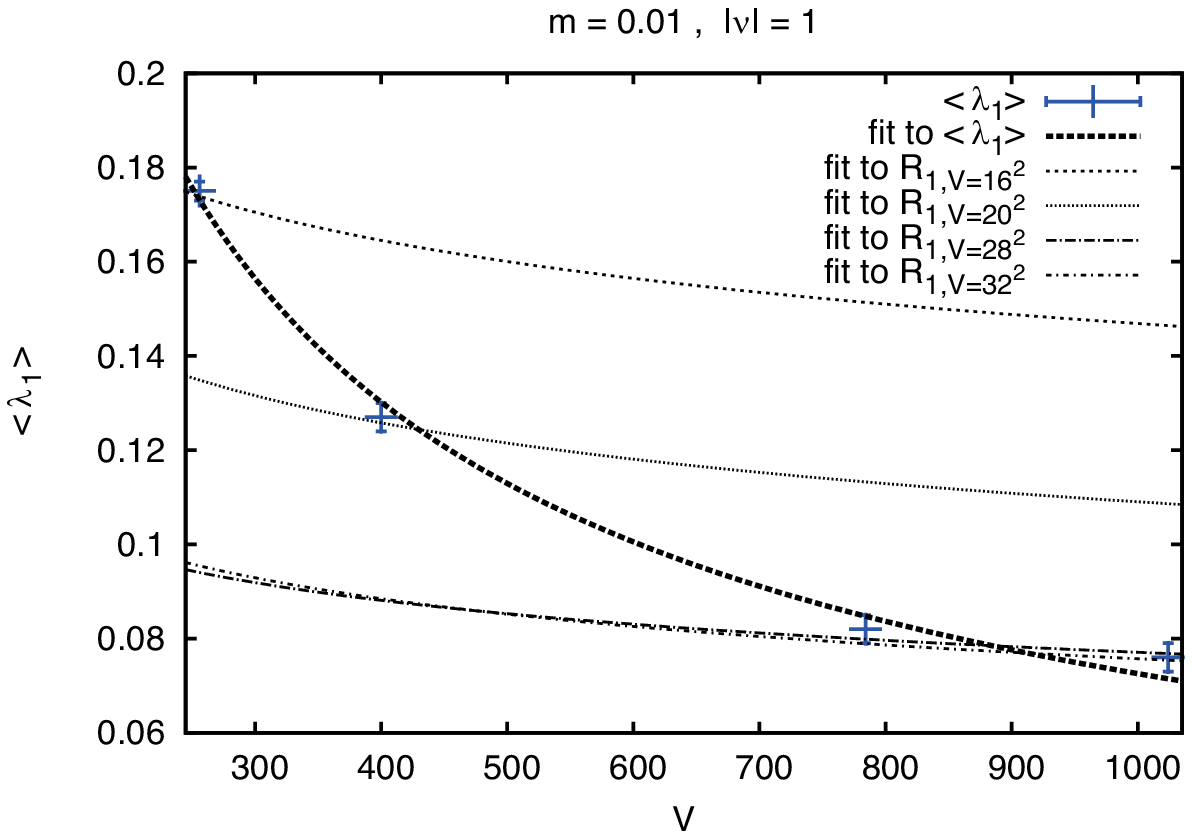}
\includegraphics[angle=0,width=.49\linewidth]{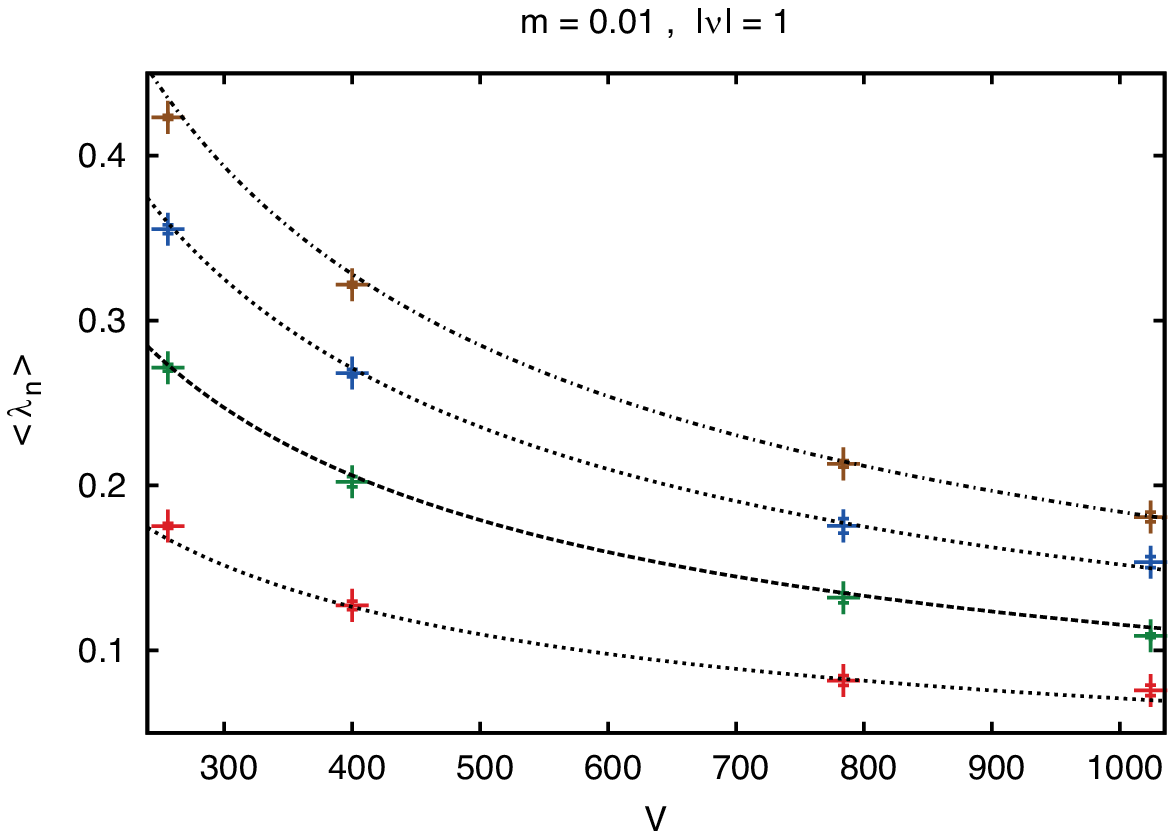}
\end{center}
\caption{The volume dependence of the mean values $\la \lambda_{n}\ra$.
The plot on the left (right) shows the data for $\la \lambda_{1}\ra$ 
(for $\la \lambda_{1}\ra \dots \la \lambda_{4}\ra \,$) in four volumes, 
which agree quite well with the $V$ dependent Kov\'{a}cs distributions
(\protect\ref{koveq}), if we fit the free parameters $\alpha$ and $c$.
However, the values of the fit parameters differ strongly from those 
required in Figure \protect\ref{cumdens3}, which adjust the detailed 
distribution in one volume. These parameter sets correspond to 
the four finer lines in the plot on the left.}
\label{meanlam}
\end{figure}

Next we consider the volume dependence of the mean values 
$\la \lambda_{n}\ra$ \cite{LMBH}. We perform fits of the
Kov\'{a}cs distribution (\ref{koveq}) to these mean 
values at $m=0.01$, in volumes $V = 16^{2} \dots 32^{2}$. 
Figure \ref{meanlam} shows that also these fits work quite well.
The plot on the left illustrates the $V$-dependence of
$\la \lambda_{1}\ra$ in the sector $|\nu |=1$; 
the fit (bold line) matches the data
in four volumes practically within the errors. The plot on the
right extends this consideration to $n=1 \dots 4$: all four 
eigenvalues in four volumes are captured well by adjusting
the two parameters $c$ and $\alpha$ in the ansatz (\ref{caform}).

So far this seems to confirm that our microscopic Dirac spectra
are compatible with the decorrelation property. However, each fit
in Figure \ref{cumdens3} was performed by tuning the two free
parameters independently. In Figure \ref{meanlam} they were
adjusted again, now to the expected volume dependence.
The consistency condition is that the values of
these parameters should be compatible, in particular the
(dimensionless) exponent $\alpha$. 

This is {\em not} the case: the $\alpha$ values 
required for these fits vary between $0.58(3)$ and $8.08(5)$ 
(and the coefficient $c$ even varies from $0.09(3)$ to
$2.7(2) \cdot 10^{4}$) \cite{LMBH}. For a graphical
illustration, the plot on the left of Figure \ref{meanlam}
also shows lines for the parameter sets which were chosen
in Figure \ref{cumdens3} for the detailed distribution
in each volume. We see that these lines are far 
from the mean values in distinct volumes.
Therefore, the agreement cannot be confirmed, although it is
amazing that the single fits work so well. \\

As a compelling test, we now focus on the distribution
of the {\em unfolded level spacings} $s$ (which is described
{\it e.g.}\ in Refs.\ \cite{BHSV,WBStanislav}). We saw in
Ref.\ \cite{BHSV} that the entire Dirac spectrum follows very
precisely the behaviour, which was predicted for the Chiral 
Unitary Ensemble (ChUE) \cite{HalVer}
\be
\rho_{\rm ChUE}(s) = 32 \Big( \frac{s}{\pi} \Big)^{2}
\exp (-4 s^{2}/\pi ) \ ,
\ee
just like QCD in the $\epsilon$-regime \cite{WBStanislav}.
On the other hand, Ref.\ \cite{KovPit} observed in high
temperature QCD with $2+1$ flavours a transition to a
Poisson distribution $\rho_{\rm Poisson}(s) = \exp (-s)$, if one
only includes low lying eigenvalues.

\begin{figure}[h!]
\center
\includegraphics[angle=0,width=.49\linewidth]{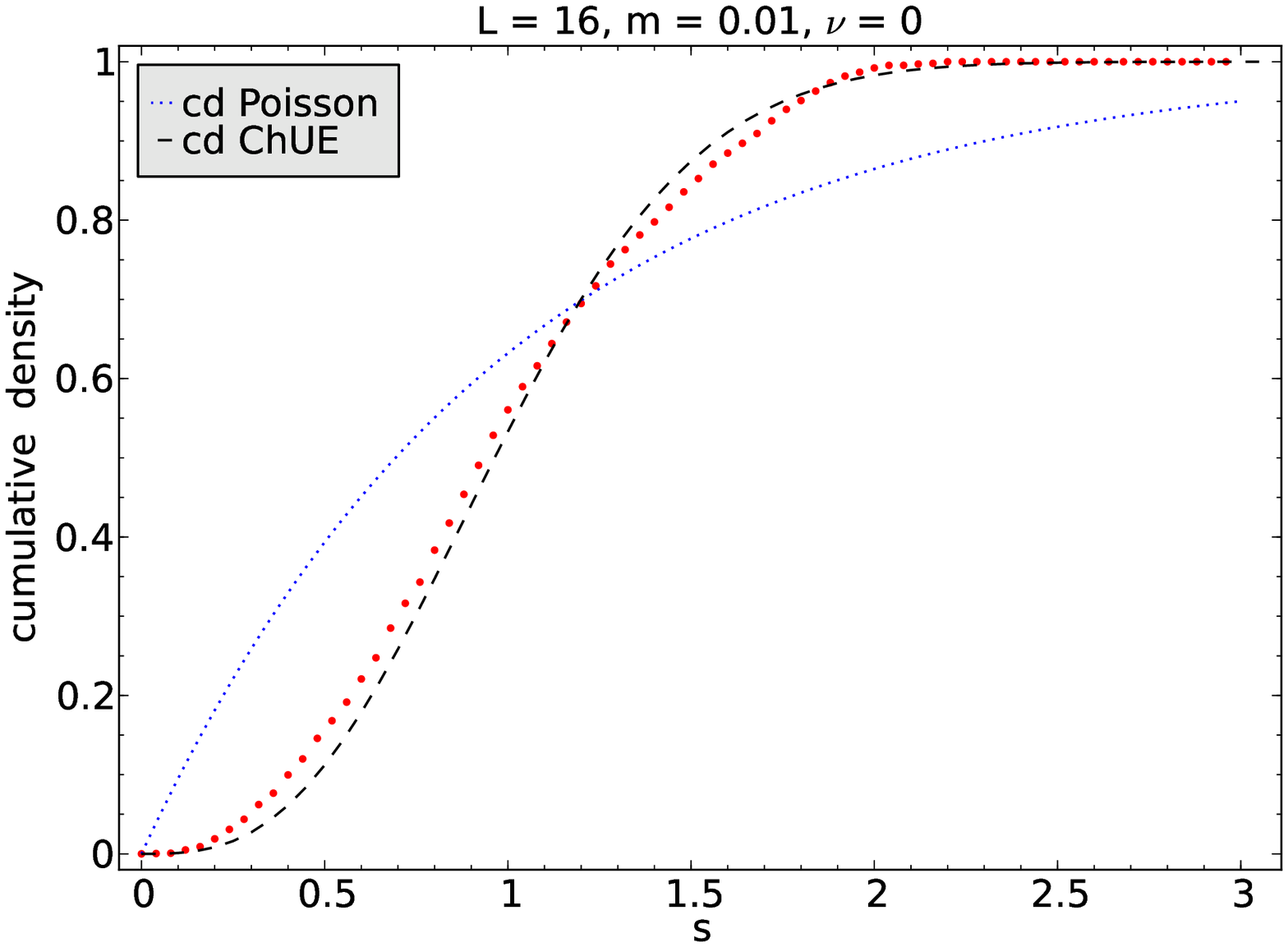}
\includegraphics[angle=0,width=.49\linewidth]{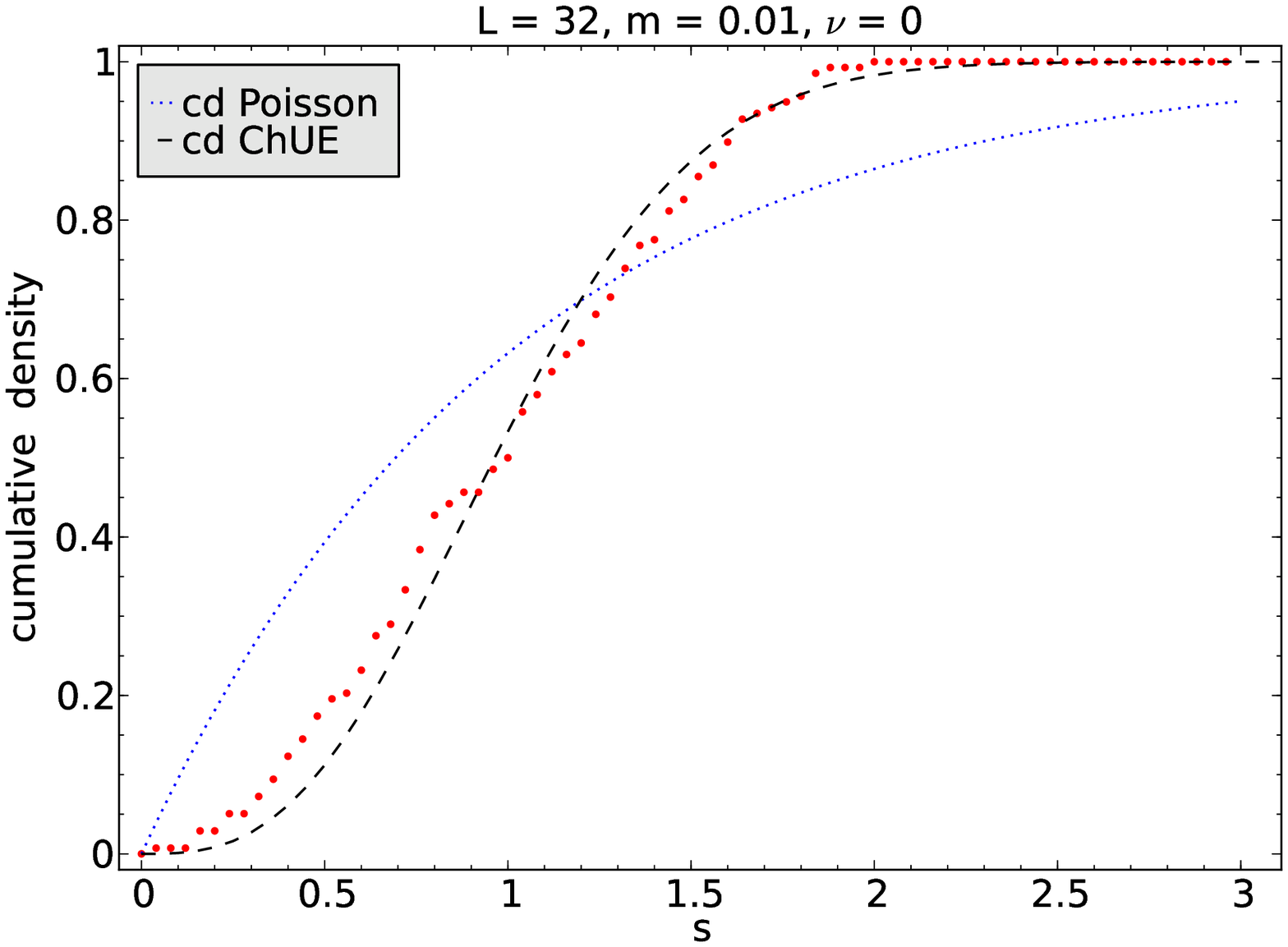}
\caption{The unfolded level spacing distributions at
 $m=0.01$, in the topological sector $\nu =0$, for the lowest two
Dirac eigenvalues at $L=16$ (on the left) and $L=32$ (on the
right). In both cases, our data are compatible with the
distribution of a Chiral Unitary Ensemble, but not with
a Poisson distribution.} 
\label{unfold}
\end{figure}

Hence we test specifically the unfolded level spacing density for 
the lowest two Dirac eigenvalues of each configuration \cite{LMBH}.
Figure \ref{unfold} shows the results for $m=0.01$, $\nu =0$
and $L=16$ (on the left), $L=32$ (on the right). In both cases
our data are close to the ChUE curve, but far from the
Poisson distribution. This is most obvious at $L=16$, where
more statistics is included (2428 configurations), while
$L = 32$ (with 138 configurations) probes smaller
eigenvalues.

\vspace*{-7mm}
\section{Mass anomalous dimension}
\vspace*{-1mm}

At last we also take a look at the evaluation of the mass anomalous 
dimension $\gamma_{m}$. This can be done in various ways; here 
we follow the procedure which was used in Ref.\ \cite{CHPS}. 
As suggested in Ref.\ \cite{Luescher}, it employs the {\em mode number}
(the cumulative eigenvalue density, up to the normalisation)
\be
\nu_{\rm mode} (\lambda ) = V \int_{-\lambda}^{\lambda} 
d \lambda ' \ \rho (\lambda ')  
= \frac{2 c V^{2}}{\alpha +1} \ \lambda ^{\alpha +1}
\ee
as a tool to evaluate the mass anomalous dimension
\be
\gamma_{m}(\lambda) = \frac{d}{\alpha (\lambda ) +1 } - 1 \ .
\ee
We are most interested in its {\em IR limit,} 
\be
\gamma_{m}^{*} = \ ^{~ \lim}_{\lambda \to 0} \ \gamma_{m} (\lambda ) \ .
\ee

\begin{figure}[h!]
\center
\includegraphics[angle=0,width=.49\linewidth]{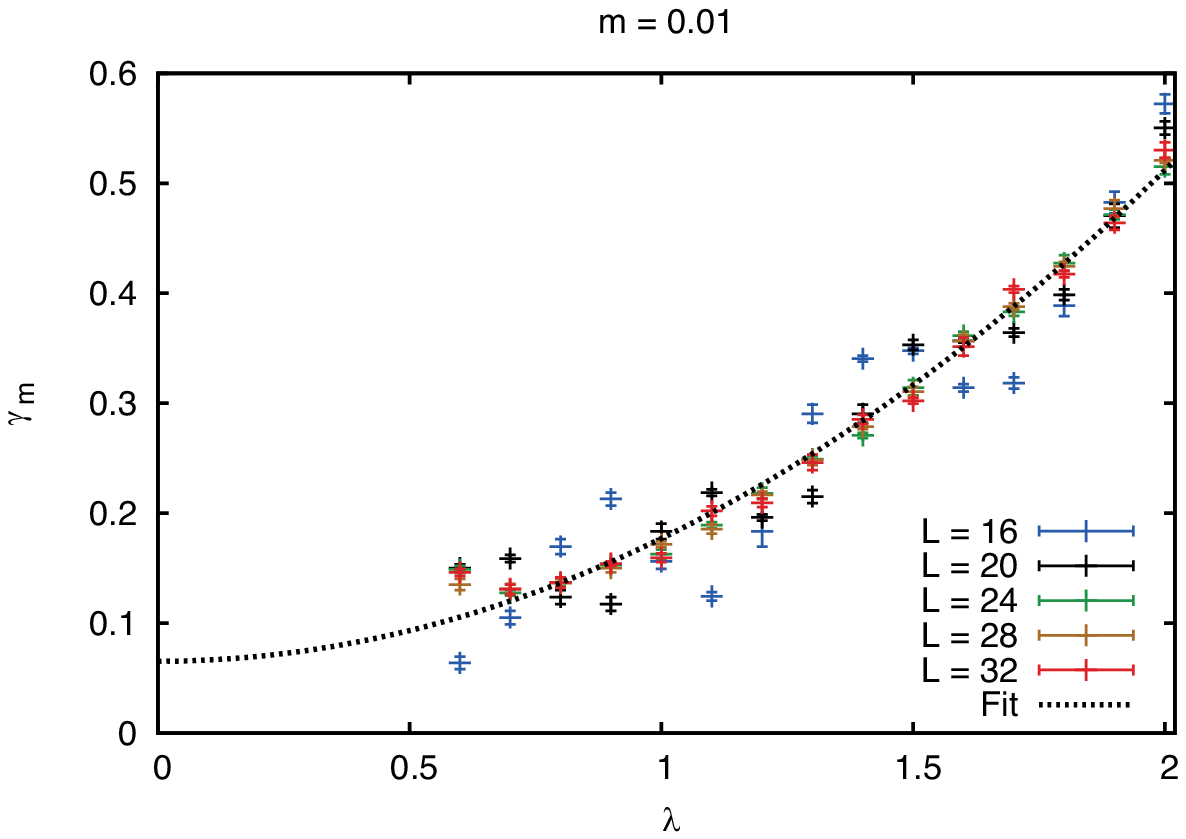}
\includegraphics[angle=0,width=.49\linewidth]{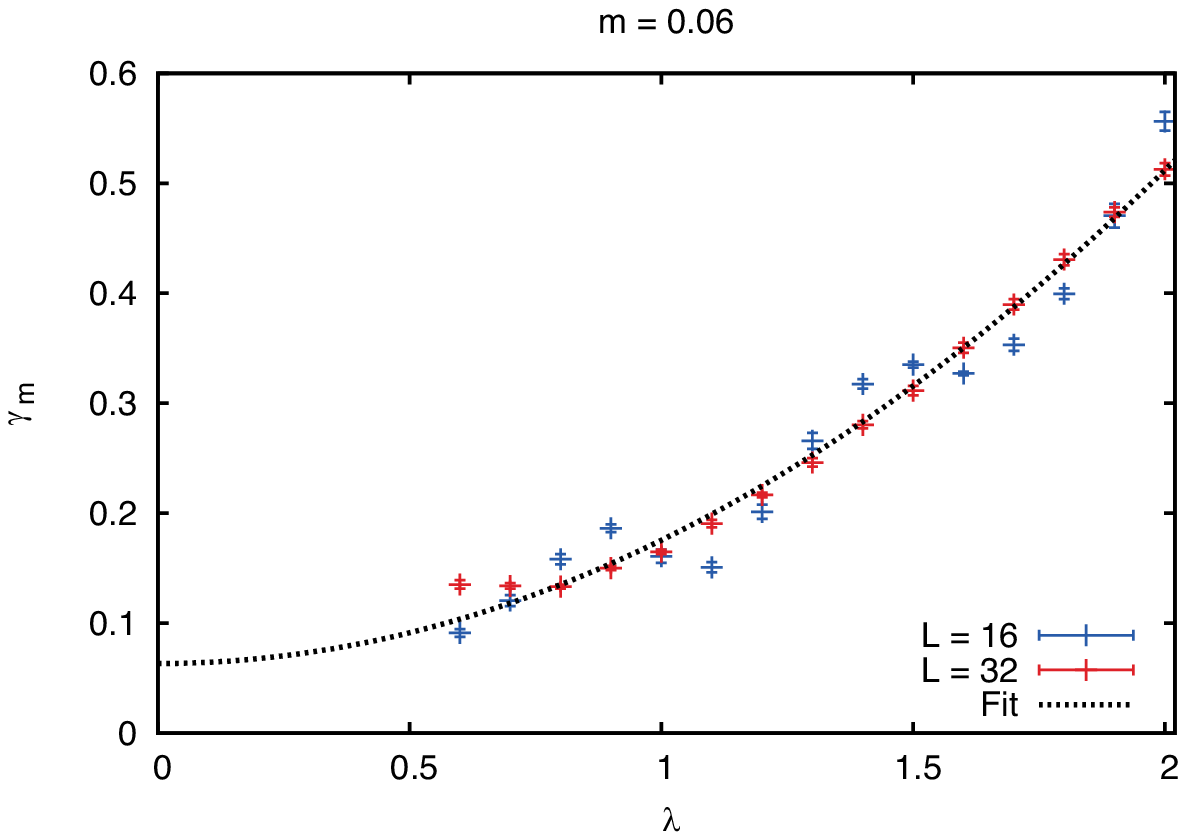}
\caption{Evaluation of the mass anomalous dimension $\gamma_{m}$
based on the mode number $\nu_{\rm mode}(\lambda )$
at fermion mass $m=0.01$ (on the left)
and $m=0.06$ (on the right). The data for both masses lead to
very similar IR limits, which are given in eq.\ 
(\protect\ref{gammastar}).}
\label{madim}
\end{figure}

Figure \ref{madim} shows the $\gamma_{m}$ values (which are obtained
by averaging over a small interval) for fermion masses $m=0.01$ and 
$0.06$, in a variety of volumes. In both cases, the data are well
compatible with a quadratic fit, which leads to very similar
IR extrapolations,
\be  \label{gammastar}
m = 0.01 \ : \ \gamma_{m}^{*} = 0.065(5) \ , \qquad
m = 0.06 \ : \ \gamma_{m}^{*} = 0.063(7) \ .
\ee
The consistency between these two masses suggests stability
of this value in the chiral extrapolation $m \to 0$, and in
the large volume extrapolation $L \to \infty$. (Moreover,
lattice artifacts are expected to be small, as we mentioned
before.)

However, this apparently convincing picture is questionable,
since the results for $\gamma_{m}^{*}$ depend significantly
on the energy interval that we rely on for the extrapolation.
Fits to the detailed distributions of the lowest few eigenvalues
only\footnote{This is similar to the method applied in 
Ref.\ \cite{lamnscaleV}.}
lead to $\alpha \simeq 3/5$ \cite{BHSV}, which corresponds
to $\gamma_{m}^{*} \simeq 1/4$. That value is just between 
the extreme limits of the HHI parameter,
\be
l \ll 1 \quad \Rightarrow \quad \gamma_{m}^{*} = 0 \ , \qquad
l \gg 1 \quad \Rightarrow \quad \gamma_{m}^{*} = 1/2 \ ,
\ee
which could be sensible regarding Table \ref{HHItab}.
In Figure \ref{madim} we refer to higher energies, so we are 
dealing with Dirac eigenvalues closer to the bulk. The
corresponding IR extrapolation approaches 
the non-anomalous value $\gamma_{m}^{*}=0$ of free fermions.

\vspace*{-1mm}
\section{Conclusions}
\vspace*{-1mm}

We have considered the 2-flavour Schwinger model as a
simple fermionic theory with a vanishing chiral condensate 
in the chiral limit, $\Sigma (m=0)=0$.
We started from the hypothesis that in such theories the
low lying Dirac eigenvalues could be decorrelated, and 
therefore Poisson distributed. In fact, this feature had been 
confirmed for fermions in $d=4$, interacting through Yang-Mills
gauge fields at high temperature \cite{Tamas,KovPit}. However,
we could not confirm this property for the 2-flavour
Schwinger model. Single fits to the predicted properties 
work remarkably well --- in particular for the density 
of a specific low eigenvalue --- but these fits require parameters,
which are clearly inconsistent. Moreover, we saw that
the unfolded level spacing distribution for the two lowest 
eigenvalues of each configuration is close to the form of the 
Chiral Unitary Ensemble, but far from a Poisson distribution;
this coincides with the feature of the entire spectrum.

These observations favour a {\em modified hypothesis:}
the predicted eigenvalue decorrelation occurs if 
$\Sigma (m=0)$ vanishes due to high temperature, but not when
this happens due to a large number of flavours.
In fact, Ref.\ \cite{BKS} associates the inverse temperature
with the localisation scale for small Dirac eigenvalues, which
is fully consistent with this modified hypothesis.

Regarding the mass anomalous dimension $\gamma_{m}$, we found an 
apparently stable IR extrapolation $\gamma_{m}^{*}$ based on 
the $\gamma_{m}$ values obtained through the mode number in a 
moderate spectral regime \cite{LMBH}. 
However, this result disagrees with fits obtained in the 
microscopic regime of the lowest Dirac eigenvalues \cite{BHSV}. 
This underscores once more that $\gamma_{m}^{*}$ is a tricky 
quantity: different methods may provide apparently convincing
values, which still differ significantly. In fact, this is 
reflected by the recent literature on possibly IR conformal
4d fermionic models 
(see {\it e.g.}\ Refs.\ \cite{CHPS,phil}), which is currently 
one of the most controversial issues in the lattice community.

\end{document}